\documentclass[aps,pra,twocolumn,showpacs,preprintnumbers,amsmath,amssymb]{revtex4-1}
\usepackage{graphicx}
\usepackage{dcolumn}
\usepackage{bm}
\usepackage{enumerate}
\usepackage{color}
\usepackage{dsfont}

\begin{document}

\title{Optical nanofiber temperature monitoring via double heterodyne detection}
\author{P. Anderson$^{1}$}
\author{S. Jalnapurkar$^{1}$}
\author{E. S. Moiseev$^{1}$}
\author{D. Chang$^{1}$}
\author{P. E. Barclay$^{1}$}
\author{A. Lezama$^{2}$}
\author{A. I. Lvovsky$^{1,3,4}$}

\affiliation{$^{1}$Institute for Quantum Science and Technology, University of Calgary, Calgary AB T2N 1N4, Canada}
\affiliation{$^{2}$Instituto de F\'isica, Facultad de Ingenier\'ia, Universidad de la Rep\'ublica,\\ J. Herrera y Reissig 565, 11300 Montevideo, Uruguay}
\affiliation{$^{3}$International Center for Quantum Optics and Quantum Technologies (Russian Quantum Center), Skolkovo, Moscow 143025, Russia}
\affiliation{$^{4}$Institute of Fundamental and Frontier Sciences, University of Electronic
	Science and Technology, Chengdu, Sichuan 610054, China.}

\email{LVOV@ucalgary.ca}
\date{\today}

\begin{abstract}
	 Tapered optical fibers (nanofibers) whose diameters are smaller than the optical wavelength are very fragile and can be easily destroyed if excessively heated by energy dissipated from the transmitted light. We present a technique for monitoring the nanofiber temperature using two-stage heterodyne detection. The phase of the heterodyne output signal is determined by that of the transmitted optical field, which, in turn, depends on the temperature through the refractive index. From the phase data, by numerically solving the heat exchange equations, the temperature distribution along the nanofiber is determined. The technique is applied to the controlled heating of the nanofiber by a laser in order to remove rubidium atoms adsorbed on its  surface that substantially degrade its transmission. Almost 90\% of the nanofiber's original transmission is recovered. 
\end{abstract}
\maketitle

\section{\label{Introduction} Introduction}
Many quantum information optical protocols require strong coupling of a single optical-field mode with individual atoms \cite{Kimble_nature}. This coupling can be enhanced by confining the optical field mode to an optical waveguide of sub-wavelength width. The interaction of the guided light with the atom can then take place through the evanescent field surrounding the waveguide. This is particularly the case for tapered optical fibers (nanofibers) whose diameter is of the order or smaller than the optical wavelength. In such case, a large fraction of the guided light energy is present in the evanescent field outside the fiber material, thereby offering ample opportunity for the coupling to free atoms \cite{dipole_trap_arnoRau}. 

Recently, there have been great strides towards reliable manufacturing of these nanofibers \cite{chormaicTNF}. The production method typically incorporates an oxy-hydrogen flame heating a portion of a commercial optical fiber. Motorized stages are then used to pull the fiber from both ends, creating a tapered section which decreases radially until a specified radius is attained \cite{chormaicTNF}. 

However, the small size and high evanescent field intensity makes the nanofibers extremely fragile. In particular, they are vulnerable to heating caused by the transmitted laser field. Although the absorption in the fiber glass itself is negligible, any imperfections or contamination of the fiber surface may absorb light and heat the fiber. Given that most nanofiber-atom coupling experiments are performed under high vacuum, the heat cannot be dissipated through convection. Heat conduction is also negligible because of the nanofiber's elongated shape. The only remaining dissipation channel, thermal radiation, becomes significant only at high temperatures of the fiber. If the temperature at which the equilibrium is reached exceeds the glass melting point, destruction of the nanofiber is likely. It is therefore important to reliably monitor the temperature of the nanofiber in the region of its waist. An additional benefit of such monitoring is that it enables controlled heating of the nanofiber, which can be useful for the desorption of contaminants from the fiber surface.

The methods suggested for nanofiber temperature monitoring rely on the temperature dependent variation of the refractive index of the fiber material (silica). This variation results in the modification of the optical path length and, consequently, the phase of the transmitted optical field. An important study of nanofiber heating was presented in Ref.~\cite{Rausch}. The optical fiber used in that work had Bragg reflecting gratings built at both ends of the fiber, thereby realizing an optical cavity. Variations of the optical phase resulted in changes of the transmission through the cavity and gave access to the temperature dependent changes of the refractive index. However, many experiments utilizing nanofibers do not require Bragg gratings. Thus, adding one or manufacturing a nanofiber from a commercial Bragg-grating optical fiber would prove arduous if only for the sake of temperature monitoring.

\section{\label{Experiment}Experiment}
In this paper we demonstrate an alternate method for nanofiber temperature monitoring. Our method exploits the phase sensitivity of balanced optical heterodyne detection and can readily be implemented with any single mode fiber without the need of Bragg reflectors at the fiber ends. The principle of the method is as follows. While the fiber is heated by a strong laser field, an additional  weak probe field is sent along the fiber. The output probe light is subjected to balanced heterodyne detection (BHD) with a local oscillator whose frequency is shifted by $\Delta$ with respect to the probe field. As a result, the BHD output is modulated at the frequency $\Delta$. A change in the signal phase results in the same change in the phase of the modulated BHD output.

The choice of the frequency $\Delta$ is usually guided by practical considerations.  It is customary to perform BHD with frequencies larger than several MHz, thereby avoiding the region where most of the technical noise in lasers occurs. In our experiment we have used an acousto-optical modulator with a resonance frequency of $\Delta = 130$ MHz. For this frequency, the BHD is shot-noise limited allowing a very good phase sensitivity. However, the time domain analysis of the signal over long time periods would require large data storage amounts. To circumvent this difficulty, we have down-shifted the oscillation frequency by performing a second heterodyne mixing of the modulated output of the BHD with a RF signal at frequency $\Delta+\delta$ with $\delta=20$ kHz. After passing through a low-pass filter (cut-off frequency $\sim$ 1 MHz) the mixer output was recorded in a digital oscilloscope and its phase variations are monitored with respect to a 20 kHz reference signal recorded on another channel of the same oscilloscope. The choice of the  frequency $\delta$ is dependent on the rate of the phase change of the probe beam. While it is desirable to minimize this frequency to reduce the data storage, it is also necessary for reliable phase monitoring that the induced change in the probe beam phase over one period ($1/\delta$) not exceed $2\pi$. 

\begin{figure}[b]

\includegraphics[width=0.95\columnwidth]{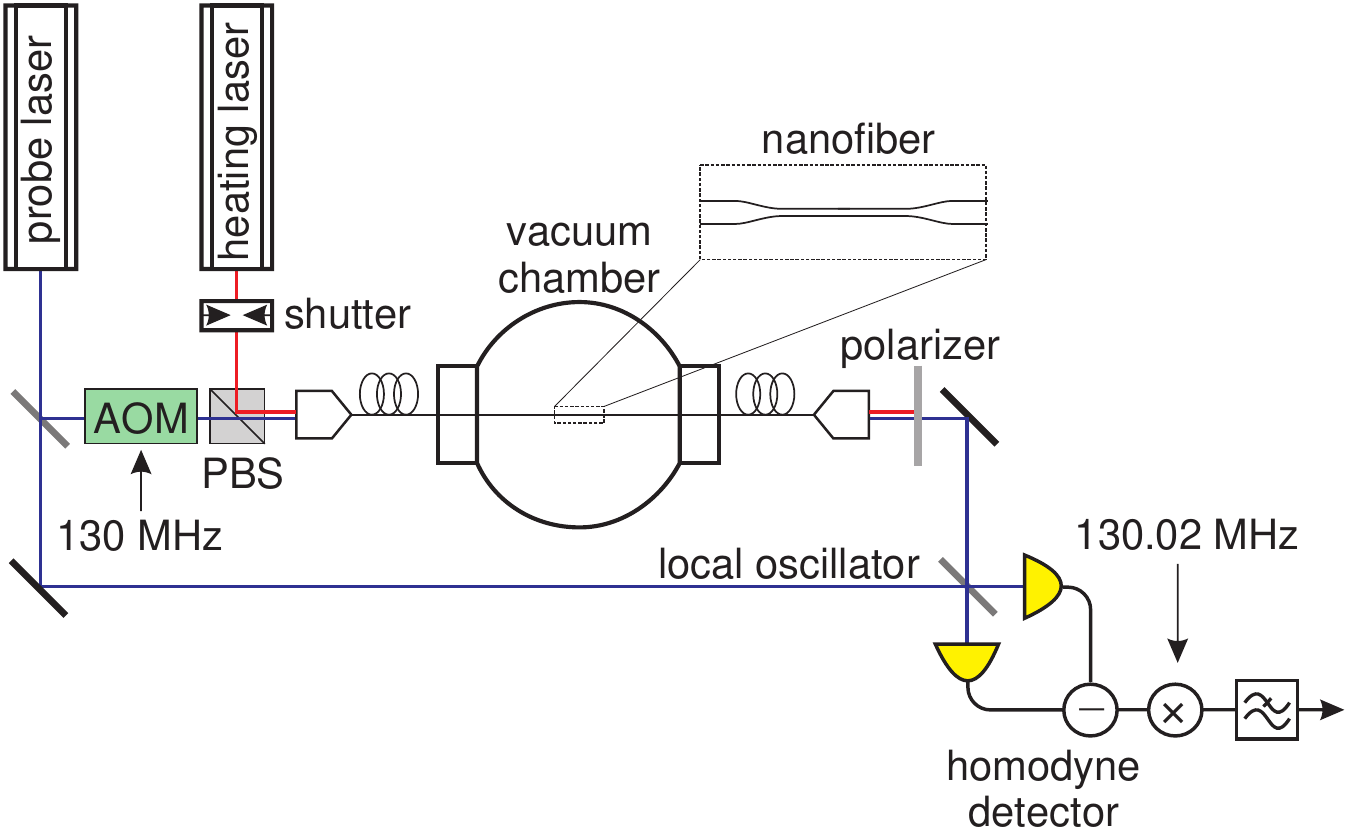}
\caption{Experimental setup. PBS: polarizing beam splitter, AOM: acousto-optical modulator.}
\label{fig:schematic}
\end{figure}

The experimental setup is shown in Fig. \ref{fig:schematic}. We have used tapered nanofiber produced from a commercial single-mode optic fiber, using an oxy-hydrogen flame and a flame-brush technique \cite{Rolston}. During the experiment, the tapered portion of the fiber was inside a vacuum chamber designed for magneto-optical cooling and trapping of rubidium. A 795 nm Ti:Sapphire ``heating'' laser was coupled into the nanofiber with a power ranging from 250 $\mu$W to 5 mW. A portion of this power is absorbed by the fiber which results in heating. The probe beam was derived from a home-made external-cavity diode laser with a wavelength $\lambda=780$ nm, allowing it to be separated from the heating beam by means of optical filters. The power of the probe beam was kept below 10 $\mu$W. After acousto-optical modulation and propagation through the fiber, the probe beam is sent to the BHD setup \cite{KUMAR20125259,Chi2011,Masalov2017} with the local oscillator obtained from the same diode laser. The two RF signals of frequencies $\Delta$ and $\Delta+\delta$ required for the double heterodyne detection were generated by the same direct digital synthesizer (DDS). The $\delta=20$ kHz phase reference signal was generated by mixing these two RF waves. 

\begin{figure}
\includegraphics[width=\columnwidth]{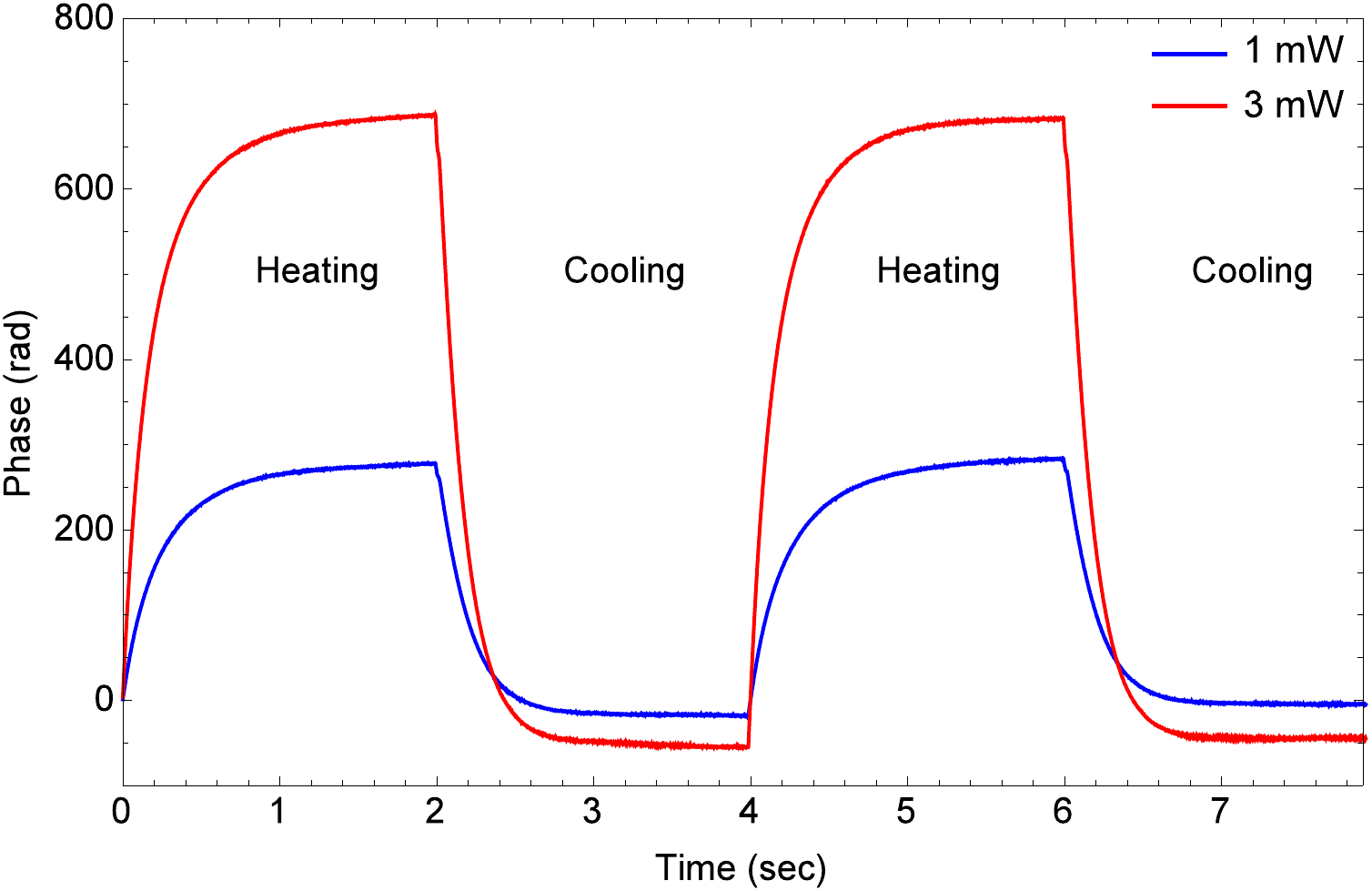}
	\caption{Phase accumulated during the heating and cooling cycles for two transmitted heating laser powers.}
    \label{fig:exp}
\end{figure}

A mechanical shutter placed on the path of the heating laser was used to create successive heating and cooling cycles of two-second duration; longer cycles produced no significant additional phase change. The measured accumulated phase variations during a heating and cooling cycle are presented in Fig.~\ref{fig:exp}, and the  measured amplitude of phase oscillations as a function of the heating power is summarized in Fig.~\ref{fig:result_plot}. In order to relate these observations to the fiber temperature, a quantitative modeling of the heat exchange in the tapered fraction of the fiber is needed. It is described in the next section.

\section{\label{Theory}Heat exchange modeling}

We describe the heat transport along the tapered portion of the fiber with the following thermodynamics equation \cite{Rausch}.
\begin{equation}\label{eq:heat}
\begin{split}
c_p\rho\partial_tTdV&=-dH_{\rm rad}(T)+dH_{\rm rad}(T_0)\\&+\lambda\partial_z^2TdV+dP_{\rm heating}+dP_{\rm gas}
\end{split}
\end{equation}
Here we neglect the variation of the fiber temperature $T$ over its cross section, accounting only for its variation along the fiber axial coordinate $z$ and time $t$. The left-hand side of Eq.~\eqref{eq:heat} represents the amount of heat per unit of time absorbed by a fragment of nanofiber of infinitesimal volume $dV$. Here, $\partial_t$ denotes the time derivative, $c_p$ the specific heat capacity of the silica, and $\rho$ its density. The first two terms on the right-hand side   describe the heat exchange by radiation: the term $-dH_{\rm rad}(T)$ represents the power radiated by the volume $d V$ while $dH_{\rm rad}(T_0)$ the absorption of thermal energy from the surrounding blackbody radiation at the room temperature $T_0$. These two terms are described by the same function of the temperature according to Kirchhoff's law. Both these terms are described by the same function of the temperature. The third term represents the heat exchange by conduction along the fiber axis $z$, with $\lambda$ being the thermal conductivity of silica. The fourth term $dP_\text{heating}$ is the heat delivered by the light propagating inside the fiber. The last term associated with the heat exchange with the environment gas can safely be neglected as the fiber is under ultra high vacuum.

\begin{figure}[h]
	\includegraphics[width=\columnwidth]{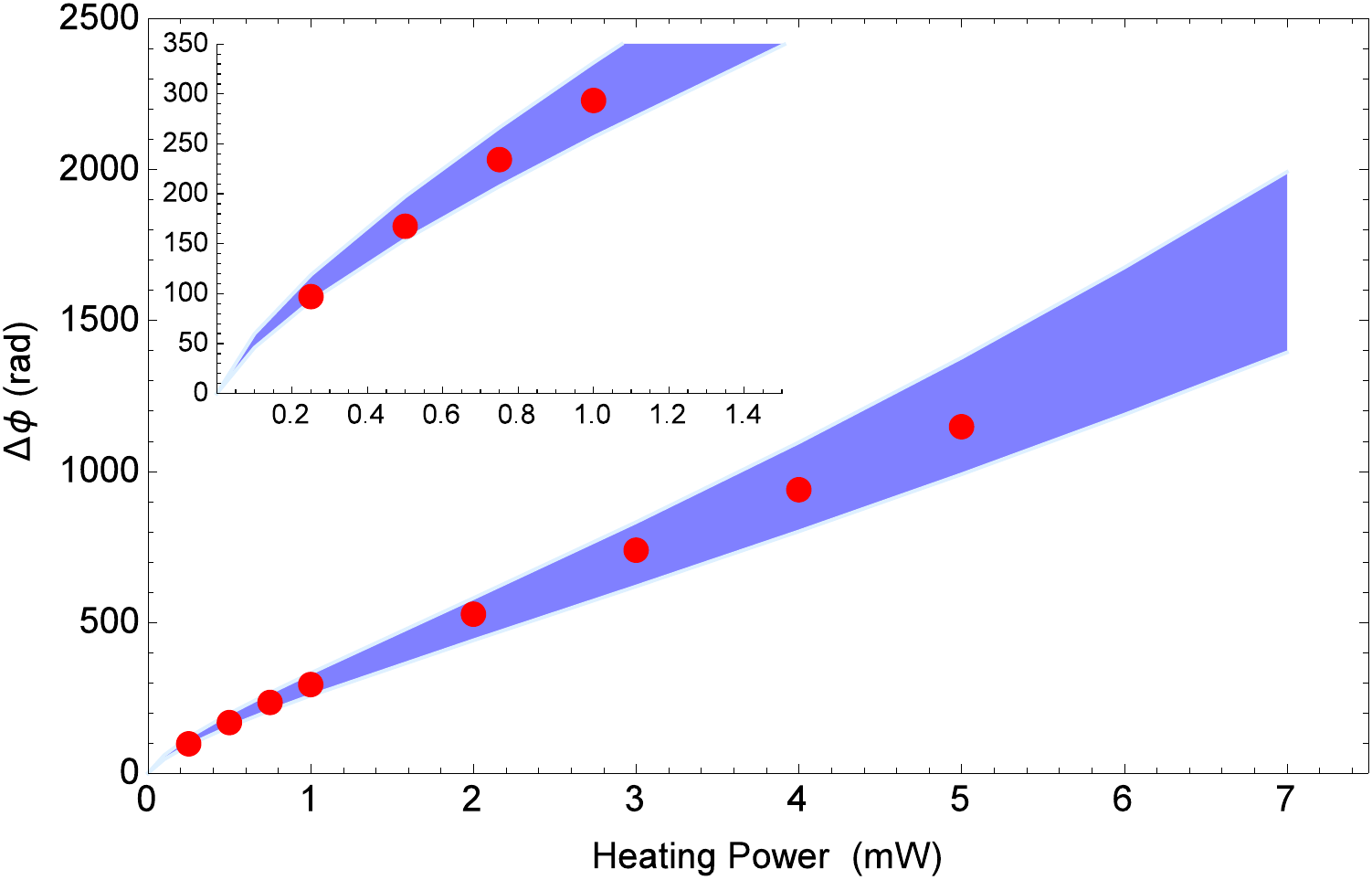}
	\caption{Phase difference of the transmitted probe light for the  the heated and cold nanofiber as a function of heating power. The experimental data (red points) are within the predicted uncertainties (blue) even for low heating powers (inset). The primary source of error is the uncertainty (5\%) in the radius profile of the nanofiber.}
	\label{fig:result_plot}
\end{figure}

\begin{figure}[h]
\includegraphics[width=\columnwidth]{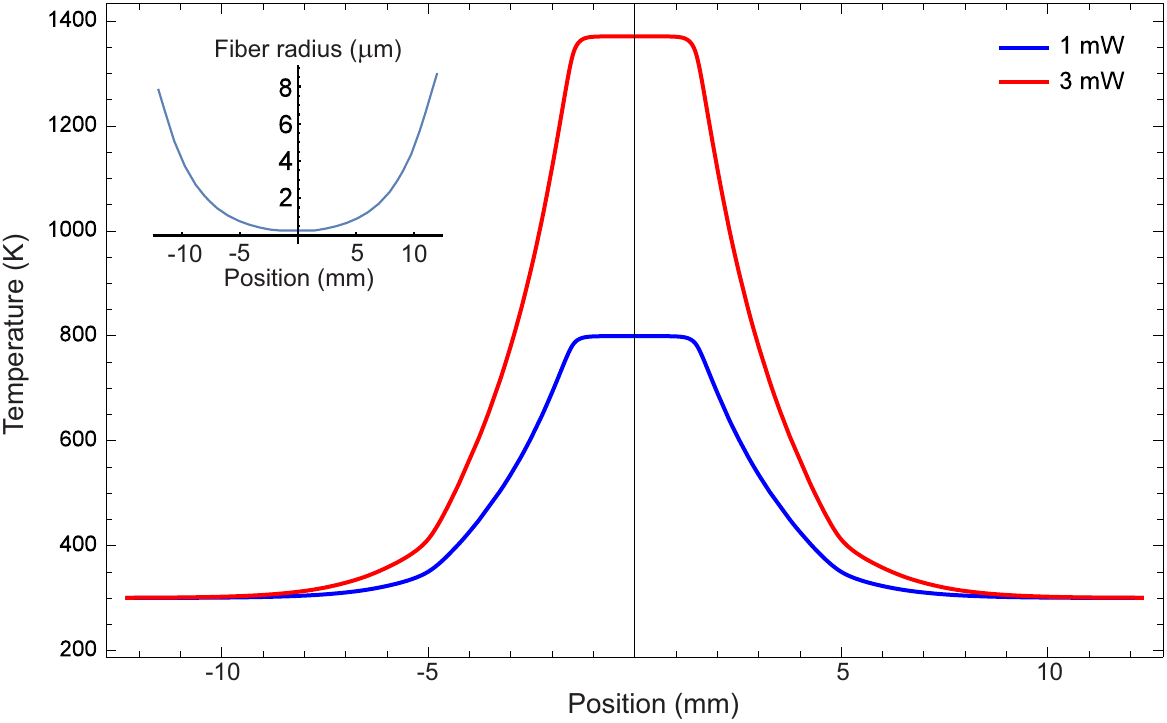}
\caption{Calculated temperature profile of the fiber at equilibrium, with the absorption parameter $k$ fit to the experimental data. The inset shows the nanofiber radius profile. }
\label{fig:temp_profile}
\end{figure}

We simplify the above equation by writing $dV=\pi a^{2}dz$, where $a=a(z)$ is the position-dependent fiber radius that is known from the fiber pulling procedure (Fig.~\ref{fig:temp_profile}, inset). Additionally, the term $dP_{\rm heating}$ can be evaluated based on the assumption that the fiber heating is essentially due to energy absorption and dissipation occurring at the surface of the fiber, predominantly due to surface pollutants \cite{fujiwara}. If we assume a constant  density of absorbers on the fiber surface, we can write
\begin{equation}\label{dpheat}
dP_\text{heating}= 2\pi a(z)k I(a)  dz 
\end{equation}
where  $k$ is a proportionality constant that depends on the surface density of absorbers and the scattering cross section, which is adjusted to fit the observed data, and $I(a)$ is the light intensity at the fiber surface. The latter quantity is readily calculated from the transverse profile of the guided mode, which is known from the literature \cite{okoshi}, and the total power of the propagating light. The details of this calculation can be found in the Appendix.

In consequence, Eq.~(\ref{eq:heat}) can be written as:
\begin{equation}\label{eq:heat_z}
\begin{split}
c_p\rho\partial_tT \pi a^{2}=-\partial_z H_{\rm rad}(T)+\partial_z H_{\rm rad}(T_0)\\
+\lambda (\partial_z^{2}T)\pi a^{2}+2\pi ak I(a)  
\end{split}
\end{equation}
This equation can be solved numerically provided that we can express the first two terms in the right-hand side --- the heat radiated from an infinitesimal fragment of the nanofiber --- as a function of its radius and temperature.
This evaluation is complicated by the fact that the thermal radiation from objects with dimensions that are comparable to or smaller than the emission wavelength are expected to considerably deviate from the Stefan-Boltzmann radiation law. In consequence, the radiated heat has to be evaluated from first principles using fluctuational electrodynamics (FED) \cite{Joulain200559}. To perform this calculation, we follow the recipe of Refs.~\cite{Rausch,wuttke}, reproducing it in the Appendix. 

To relate the computed temperature variation to the experimental observations we consider the change in the optical path-length $\Delta{l}$:
\begin{equation}\label{eq:length}
\Delta{l}=\int_{-l_{0}/2}^{l_{0}/2}( n_{\rm eff}(T,z)-n_{\rm eff}(T_{0},z))dz 
\end{equation}
Here $n_{\rm eff}(T)$ is the temperature dependent effective refractive index, $l_0$ is the length of fiber assumed to be a constant (the effect of the fiber thermal expansion is negligible \cite{wuttke}).  The change in the probe field phase is then given by $\Delta{\phi}=\frac{2\pi\Delta{l}}{\lambda}$.

This calculation results in the theoretical prediction of the probe phase shift as a function of the heating power that depends on a single fit parameter, $k$. As is evident from Fig.~\ref{fig:result_plot}, a good agreement with the experimental data is present.  The parameter $k$ extracted from the fit provides important information about the surface density of contaminants, which can be used to compare nanofibers obtained in different settings. 

Figure \ref{fig:temp_profile} shows the calculated temperature variation along the nanofiber. Notice the plateau around the fiber minimum waist. The small variation of the temperature in this region is a consequence of the almost constant diameter of the fiber and negligible thermal conductivity.

\section{\label{Application}Rubidium desorption}

Our experimental apparatus is intended for the efficient coupling of magneto-optically trapped $^{^{87}}$Rb atoms with the evanescent optical field surrounding the nanofiber. One challenge faced by our system is the accumulation of atoms on the surface of the fiber. As atoms adsorb onto the fiber, its transmission decreases due to the scattering of the coupled light \cite{fujiwara}. This is detrimental not only to the quantum optical properties of the transmitted light, but also to the structural stability of the fiber since the scattered light leads to localized heating. A solution that is frequently employed is illuminating the inside of the vacuum chamber with ultraviolet light. This approach is however unsuitable if the windows of the vacuum chamber do not transmit ultraviolet radiation. Under these conditions, an alternative option consists in controllably heating the fiber, enough to desorb the atoms, but without melting it. 

Lai {\it et al.}~investigated the possibility of recovering, and preserving, the transmission of a nanofiber in rubidium vapor by means of  an external heater mounted onto the fiber holder \cite{Lai:13}. This enabled them to recover a portion of the lost transmission, however the recovery was incomplete. Also, an external heating apparatus considerably reduces the optical access to the nanofiber, complicating its coupling to ultracold atoms in an optically trapped cloud.  

\begin{figure}[t]
	\includegraphics[width=0.8\columnwidth]{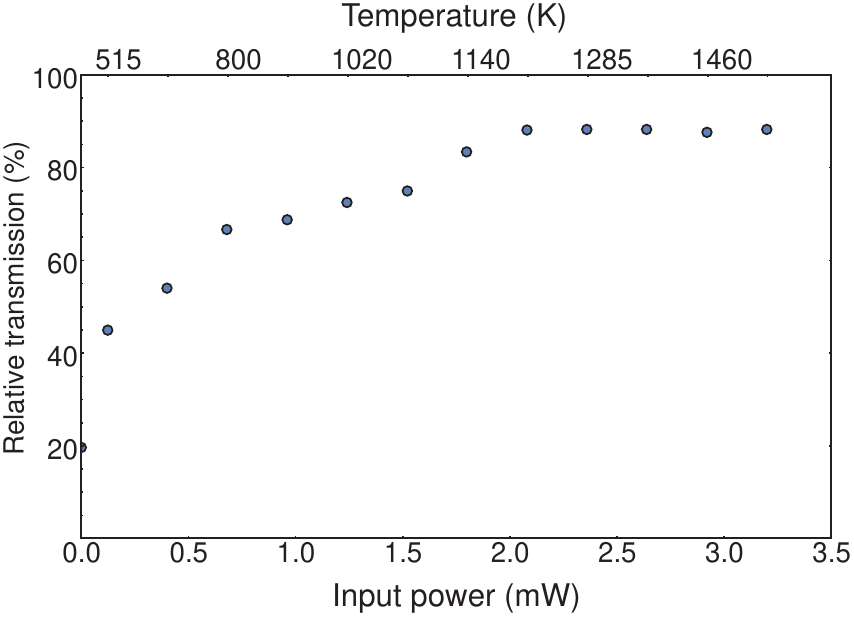}
	\caption{Relative transmission of the fiber as a function of the heating laser power.}
	\label{fig:recoveryfig}
\end{figure}

Making use of our ability to precisely monitor the nanofiber temperature, we have studied the desorption of Rb atoms from the fiber surface as it is heated by the heating laser. Initially, a cold atomic cloud of $^{87}$Rb was formed in a magneto-optical trap in the proximity of the nanofiber.  After overlapping the rubidium cloud with the fiber waist, we observed a significant drop (more than 80\%) in the transmission of the probe field. Then we applied the heating laser, whose  power was  slowly increased in 25 $\mu$W increments. The results of this experiment are reported in Fig \ref{fig:recoveryfig}. A recovery of the fiber transmission to near 90\% of the original value was observed when the temperature reached about 1200 K.
 
\section{\label{Conclusions}Conclusion}
In summary, we have successfully implemented the thermometry of tapered optical nanofibers using double heterodyne detection. We further applied controlled heating to the recovery of fiber transmission after antecedent loss due to immersion in a cold Rubidium cloud. The ability to recover and preserve the transmission of a tapered nanofiber is an important tool for cold atom experiments. The ability to do this without any added equipment or changes to the fiber serves as an even greater utility.

\appendix
\section{Radiated power}
The following is a model calculating the heat radiation of a cylinder reproduced from \cite{golyk}. The heat radiated per unit length of an infinite cylinder made of an isotropic material, with radius $a$ at temperature $T_c$,  is 
\begin{equation}\label{eq:dzH}
\begin{split}
\partial_z{H_{\rm rad}}
&=\frac{4}{c_0}\int_0^\infty{d\nu}\frac{h\nu^2}{\text{exp}(h\nu/k_BT_c)-1}\\ \times\sum_{P=\bot,\parallel}&\sum_{l=-\infty}^{\infty}\int_{-1}^{1}d\xi(\text{Re}(\mathds{T}_{l,\xi}^{PP})+|\mathds{T}_{l,\xi}^{PP}|^2+|\mathds{T}_{l,\xi}^{P\overline{P}}|^2),
\end{split}
\end{equation}
where $c_0$ is the speed of light in vacuum, $\nu$ is the optical frequency, $P$ is the polarization of the radiated light with respect to the plane of propagation, $l$ is the mode number and $\overline{P}$ denotes the perpendicular polarization to $P$. The quantity $\xi=\frac{k_z}{k_0}$, where $k_0=2\pi\nu/c_0$ is the vacuum propagation constant, is the ratio between the axial and vacuum propagation constants. 
The matrix $\mathds{T}$ represents the cylindrical geometry of the emitter; its elements are as follows \cite{golyk}
\begin{equation}\label{eq:Tmatrix}
\begin{split}
\mathds{T}^{\bot\bot}_{l,\xi}&=-\frac{J_l(qa)}{H^{(1)}_l(qa)}\frac{\Delta_1\Delta_4-K^2}{\Delta_1\Delta_2-K^2},\\
\mathds{T}^{\parallel\parallel}_{l,\xi}&=-\frac{J_l(qa)}{H^{(1)}_l(qa)}\frac{\Delta_2\Delta_3-K^2}{\Delta_1\Delta_2-K^2},\\
\mathds{T}^{\bot\parallel}_{l,\xi}=T^{\parallel\bot}_{l,\xi}&=\frac{2iK}{\pi\sqrt{\epsilon\mu}(qaH^{(1)}_l(qa))^2}\frac{1}{\Delta_1\Delta_2-K^2},\\
\end{split}
\end{equation}
where 
\begin{equation}\label{eq:deltas}
\begin{split}
\Delta_1&=\frac{J'_l(q_1a)}{q_1aJ_l(q_1a)}-\frac{1}{\epsilon}\frac{H^{(1)\prime}_l(qa)}{qaH^{(1)}_l(qa)},\\
\Delta_2&=\frac{J'_l(q_1a)}{q_1aJ_l(q_1a)}-\frac{1}{\mu}\frac{H^{(1)\prime}_l(qa)}{qaH^{(1)}_l(qa)},\\
\Delta_3&=\frac{J'_l(q_1a)}{q_1aJ_l(q_1a)}-\frac{1}{\epsilon}\frac{J^{\prime}_l(qa)}{qaJ_l(qa)},\\
\Delta_4&=\frac{J'_l(q_1a)}{q_1aJ_l(q_1a)}-\frac{1}{\mu}\frac{J^{\prime}_l(qa)}{qaJ_l(qa)},\\
K&=\frac{l\xi{k_0}c_0}{\sqrt{\epsilon\mu}{a^2}\omega}\left(\frac{1}{q^2_1}-\frac{1}{q^2}\right),\\
q&=k_0\sqrt{1-\xi^2}, \text{ and}\\
q_1&=k_0\sqrt{\epsilon\mu-\xi^2}\\
\end{split}
\end{equation}
$J_l(x)$ is the Bessel function of the first kind, $H^{(1)}_l(x)$ is the Hankel function of the first kind, $\epsilon$ and $\mu$ represent the relative permittivity and permeability of silica and the prime indicates a derivative such that $f'(x)=\partial_xf(x)$. 

\section{Field intensity at fiber surface}
The intensity of the heating laser field at the nanofiber surface is given by the Poynting vector, which is obtained from the electric and magnetic fields of the mode propagating through the fiber. The equations for these fields  are available in literature (e.g.~\cite{okoshi}); we reproduce them here for reference. $HE_{11}$ is the only mode that needs to be considered as only this mode can be guided through the tapered region of the fiber. 

Inside the fiber core ($r \leq a$), the axial ($z$), radial ($r$) and circumferential ($\theta$) components of the fields at the point with the cylindrical coordinates $(r,\theta)$ with respect to the fiber axis are given by

\begin{align}\label{eq:Einside}
E_z&=AJ_1\left(\frac{ur}{a}\right)\sin(\theta)\\ \nonumber
E_r&=\left[-A\frac{i\beta}{u/a}J'_1\left(\frac{ur}{a}\right)+B\frac{i\omega\mu_0}{(u/a)^2}\frac{1}{r}J_1\left(\frac{ur}{a}\right)\right]\sin(\theta)\\ \nonumber
E_\theta&=\left[-A\frac{i\beta}{(u/a)^2}\frac{1}{r}J_1\left(\frac{ur}{a}\right)+B\frac{i\omega\mu_0}{u/a}J'_1\left(\frac{ur}{a}\right)\right]\cos(\theta);
\end{align}
\begin{align}\label{eq:Hinside}
H_z&=BJ_1\left(\frac{ur}{a}\right)\cos(\theta)\\ \nonumber
H_r&=\left[A\frac{i\omega\epsilon}{(u/a)^2}\frac{1}{r}J_1\left(\frac{ur}{a}\right)-B\frac{i\beta}{u/a}J'_1\left(\frac{ur}{a}\right)\right]\cos(\theta)\\ \nonumber
H_\theta&=\left[-A\frac{i\omega\epsilon}{u/a}J'_1\left(\frac{ur}{a}\right)+B\frac{i\beta}{(u/a)^2}\frac{1}{r}J_1\left(\frac{ur}{a}\right)\right]\sin(\theta),
\end{align}
where
\begin{equation}\label{eq:norm_prop_const}
\begin{split}
u=k_0a\sqrt{n^2_1-\beta^2_{11}}.
\end{split}
\end{equation}
For $r > a$, the field components are as follows:
\begin{align}\label{eq:Eoutside}
E_z&=CK_1\left(\frac{wr}{a}\right)\sin(\theta)\\ \nonumber
E_r&=\left[C\frac{i\beta}{w/a}K'_1\left(\frac{wr}{a}\right)-D\frac{i\omega\mu_0}{(w/a)^2}\frac{1}{r}K_1\left(\frac{wr}{a}\right)\right]\sin(\theta)\\ \nonumber
E_\theta&=\left[C\frac{i\beta}{(w/a)^2}\frac{1}{r}K_1\left(\frac{wr}{a}\right)-D\frac{i\omega\mu_0}{w/a}K'_1\left(\frac{wr}{a}\right)\right]\cos(\theta);
\end{align}
\begin{align}\label{eq:Houtside}
H_z&=DK_1\left(\frac{wr}{a}\right)\cos(\theta)\\ \nonumber
H_r&=\left[-C\frac{i\omega}{(w/a)^2}\frac{1}{r}K_1\left(\frac{wr}{a}\right)+D\frac{i\beta}{w/a}K'_1\left(\frac{wr}{a}\right)\right]\cos(\theta)\\ \nonumber
H_\theta&=\left[C\frac{i\omega}{w/a}K'_1\left(\frac{wr}{a}\right)-D\frac{i\beta}{(w/a)^2}\frac{1}{r}K_1\left(\frac{wr}{a}\right)\right]\sin(\theta).
\end{align}
where
\begin{equation}\label{eq:norm_atten_const}
\begin{split}
w=k_0a\sqrt{\beta^2_{11}-n^2_2}.
\end{split}
\end{equation}
In the above equations, $k_0$ is the propagation constant of the light field in the vacuum, $\beta_{11}$ the propagation constant for the $HE_{11}$ mode in the fiber, $\theta$ is the angle of the incoming linearly polarized light, $\mu_0$ is the magnetic constant and $\epsilon$ is the electric permittivity of silica. The light entering the fiber is assumed linearly polarized along $\theta=0$.

The coefficients $A, B, C$ and $D$ must satisfy the boundary conditions for the electric and magnetic fields inside and outside the fiber, which results in the following relations:
\begin{equation}\label{eq:coeff}
\begin{split}
&AJ_1(u)-CK_1(w)=0\\
&BJ_1(u)-DK_1(w)=0\\
&A\frac{i\beta}{(u/a)^2}\frac{J_1(u)}{a}-B\frac{i\omega\mu_0}{u/a}J'_1(u)\\&+C\frac{i\beta}{(w/a)^2}\frac{K_1(w)}{a}-D\frac{i\omega\mu_0}{w/a}K'_1(w)=0\\
&A\frac{i\omega\epsilon_1}{u/a}J'_1(u)-B\frac{i\beta}{(u/a)^2}\frac{J_1(u)}{a}\\&+C\frac{i\omega\epsilon_2}{w/a}K'_1(w)-D\frac{i\beta}{(w/a)^2}\frac{K_1(w)}{a}=0\\
\end{split}
\end{equation}

Finally, we can obtain the surface intensity from Eqs.~\eqref{eq:Einside} and \eqref{eq:Hinside} by calculating the axial component of the Poynting vector:
\begin{equation}\label{eq:poynting}
\begin{split}
I(a)=S_z(a)=\left.E_r{H_\theta}-E_\theta H_r\right|_{r=a}.
\end{split}
\end{equation}

\bibliographystyle{apsrev}
\bibliography{fiberpaper}

\end{document}